\documentclass[aps,prb,twocolumn,floatfix,10pt,superscriptaddress]{revtex4-1}
\usepackage[utf8]{inputenc}

\usepackage{natbib}
\usepackage{graphicx}
\usepackage{latexsym}
\usepackage{amssymb}
\usepackage{amsmath}
\usepackage{amsfonts}
\usepackage{gensymb}
\usepackage{bm}
\usepackage{hyperref}
\usepackage{braket}
\usepackage{tikz}
\usepackage{rotating}
\usetikzlibrary{arrows}
\usetikzlibrary{decorations.pathreplacing,decorations.markings}
\usepackage[normalem]{ulem} 

\newcommand{\refcite}[1]{Ref.~\onlinecite{#1}}
\newcommand{\eqnref}[1]{Eq.~(\ref{#1})}
\newcommand{\figref}[1]{Fig.~\ref{#1}}
\newcommand{\sfigref}[2]{Fig.~\hyperref[#1]{\ref{#1}(#2)}}

\newcommand{\secref}[1]{Sec.~\ref{#1}}

\hypersetup{
  colorlinks = true,
  citecolor = blue,
  urlcolor= blue,
  pdfauthor = {Wilbur Shirley, Kevin Slagle, Xie Chen},
  pdftitle = {Shirley et al. - 2018 - Foliated fracton order in the checkerboard model}
}

\begin{document}

\title{Foliated fracton order in the checkerboard model}
\date{\today}

\author{Wilbur Shirley}
\affiliation{Department of Physics and Institute for Quantum Information and Matter, California Institute of Technology, Pasadena, California 91125, USA}

\author{Kevin Slagle}
\affiliation{Department of Physics, University of Toronto, Toronto, Ontario M5S 1A7, Canada}
\affiliation{Department of Physics and Institute for Quantum Information and Matter, California Institute of Technology, Pasadena, California 91125, USA}

\author{Xie Chen}
\affiliation{Department of Physics and Institute for Quantum Information and Matter, California Institute of Technology, Pasadena, California 91125, USA}

\begin{abstract}

In this work, we show that the checkerboard model exhibits the phenomenon of foliated fracton order. We introduce a renormalization group transformation for the model that utilizes toric code bilayers as an entanglement resource, and show how to extend the model to general three-dimensional manifolds. Furthermore, we use universal properties distilled from the structure of fractional excitations and ground-state entanglement to characterize the foliated fracton phase and find that it is the same as two copies of the X-cube model. Indeed, we demonstrate that the checkerboard model can be transformed into two copies of the X-cube model via an adiabatic deformation.
\end{abstract}

\maketitle

\section{Introduction}

Fracton models
\cite{VijayFracton,Sagar16,FractonRev,3manifolds,Slagle17Lattices,CageNet,VijayNonabelian,HaahCode,MaLayers,ChamonModel,ChamonModel2,YoshidaFractal,HsiehPartons,HalaszSpinChains,VijayCL,Slagle2spin,FractalSPT,YouSondhiTwisted,SongTwistedFracton,Petrova_Regnault_2017}
are a collection of gapped three-dimensional lattice models that share a range of exotic properties.
\cite{FractonStatistics,DevakulCorrelationFunc,KubicaYoshidaUngauging,WilliamsonUngauging,Slagle17QFT,BulmashFractal,Schmitz_2017,YouSSPT,Bravyi11,PremGlassy,PinchPoint,DevakulWilliamson}
Most saliently, they contain quasiparticle excitations with constrained mobility and exhibit a ground state degeneracy that scales exponentially with linear system size.\cite{VijayFracton,ChamonModel2} Moreover, the entanglement entropy of a region contains a sub-leading correction to the area law that is proportional to the diameter of the region.\cite{FractonEntanglement,ShiEntropy,HermeleEntropy,BernevigEntropy} At the same time, each model appears to differ drastically from other models. Most strikingly, some fracton models contain string-like operators as logical operators on the ground space while others do not. \cite{YoshidaFractal,HaahCode} Furthermore, the quasiparticle content in varying models differ in number, allowed movement pattern, and statistics.\cite{FractonStatistics} The scaling constants in the ground state degeneracy and entanglement entropy vary between models as well.

A natural question to ask is whether the `fracton order' in various models is the same or different. In other words, we want to know whether the differences between a given pair of models are merely superficial or if they reflect a fundamental distinction between the two models in terms of their universal properties. This question has been difficult to answer in the absence of a clear definition of `fracton order' and a clear distinction between universal and non-universal properties of fracton models.

In \refcite{3manifolds}, we addressed this question by presenting an explicit definition of the so-called \textit{foliated fracton phases} (FFP), which covers a large subset of all fracton models.\footnote{
Gapless $U(1)$ fracton models \cite{PretkoU1,electromagnetismPretko,Rasmussen2016,Xu2006,PretkoTheta,PretkoDuality,PretkoGravity,GromovElasticity,MaPretko18,BulmashHiggs,MaHiggs,PaiFractonicLines} and type-II fracton models (in which excitations are created at corners of fractal operators) \cite{HaahCode,YoshidaFractal} are not captured by the notion of foliated fracton phases.}
Based on this definition, in Refs. ~\onlinecite{FractonEntanglement} and ~\onlinecite{FractonStatistics} we discussed universal properties of FFPs pertaining to their entanglement entropy and fractional excitation types and statistics. Consideration of these properties subsequently enables us to compare the foliated fracton order in different models.

The basic idea behind the definition of FFP is that we are concerned only with the non-trivial behavior intrinsic to three dimensions, and hence we should `mod out' the topological behavior arising from the 2D layers of the underlying foliation structure. That is, when determining the FFP equivalence relation between 3D fracton models, 2D models should be considered as free resources. Thus, two 3D models are considered as equivalent if they can be smoothly connected after the addition of gapped 2D layers. This drastically changes the usual notion of gapped topological phase as two models in the same FFP can have different ground state degeneracy and different numbers of fractional excitations since the 2D resources can carry non-trivial ground state degeneracy and fractional excitations themselves. By modding out features coming from 2D layers, the universal properties of the foliated fracton models can be characterized by a much simpler and robust set of data which can then be compared between models.

In particular, we demonstrated in \refcite{3manifolds} that the X-cube model\cite{Sagar16} belongs to a FFP. Its universal properties can be analyzed as discussed in Refs.~\onlinecite{FractonEntanglement, FractonStatistics}. In fact, we showed that the X-cube model is a renormalization group fixed point in the FFP as the system size can be increased (or decreased) by adding (or removing) layers of 2D toric codes and applying local unitary transformations. In this paper, we show that the checkerboard model\cite{Sagar16} is also a fixed point of a FFP. By comparing the universal properties
of the X-cube and checkerboard models and by establishing carefully an exact mapping, we actually show that the checkerboard model is equivalent to two copies of the X-cube model up to a generalized local unitary transformation.\cite{Xie10}

The paper is organized as follows: In section~\ref{sec:model}, we briefly review the definition of the model and some simple properties. In section~\ref{sec:RG}, the RG transformation for the model is presented which utilizes 2D toric code bilayers as resources. In section~\ref{sec:manifolds}, we show that the model can be defined on general three-manifolds equipped with a total foliation structure and derive the general formula for ground state degeneracy. In section~\ref{sec:entropy}, entanglement entropy in the ground state wave function is studied using the scheme proposed in Ref.~\onlinecite{FractonEntanglement}. In section~\ref{sec:excitations}, the fractional excitations of the model are studied using the framework developed in Ref.~\onlinecite{FractonStatistics}. This analysis collectively points to the fact that the checkerboard model is equivalent to two copies of the X-cube model as a foliated fracton phase. We present an explicit mapping between the two in section~\ref{sec:mapping}. Finally we conclude with a brief discussion in section~\ref{sec:discussion}.

\section{The checkerboard model}

\label{sec:model}

\begin{figure}
    \centering
    \includegraphics[width=6cm]{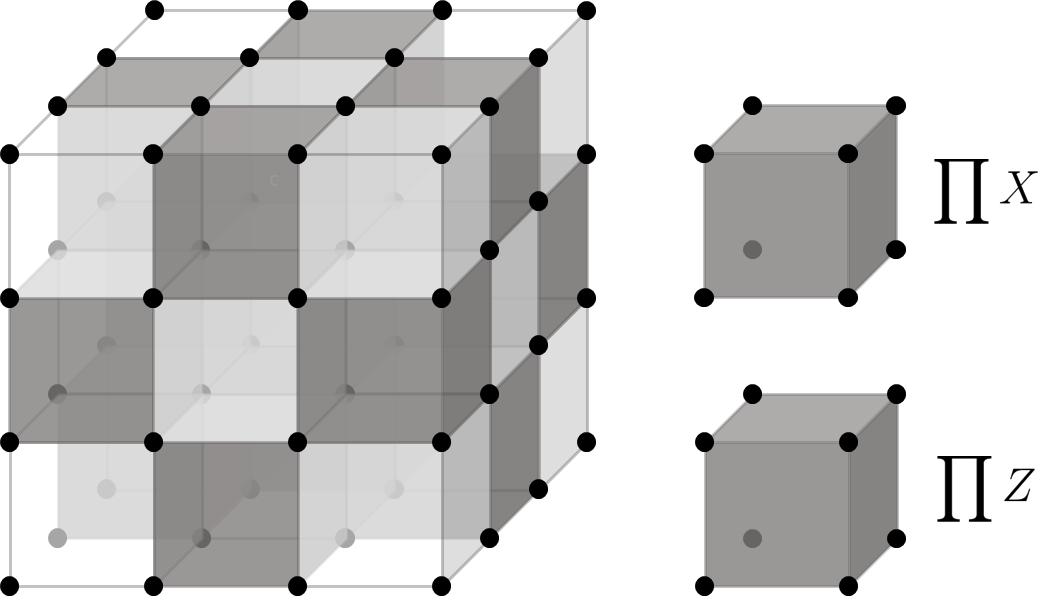}
    \caption{(a) $A$-$B$ checkerboard bipartition of cubic lattice cells. The darkened cells belong to the $A$ sublattice. Black dots represent qubits. (b) $X_c$ and $Z_c$ Hamiltonian terms. Here, $\prod X$ ($\prod Z$) denotes a product of $X$ ($Z$) operators over the depicted qubits.}
    \label{fig:model}
\end{figure}

The checkerboard model, as first discussed in \refcite{Sagar16}, is defined on a cubic lattice with one qubit degree of freedom per vertex. The elementary cubes of the lattice are bipartitioned into $A$ and $B$ 3D checkerboard sublattices, and the Hamiltonian is defined as follows:
\begin{equation}
    H=-\sum_{c\in A}X_c-\sum_{c\in A}Z_c,
    \label{eqn:H}
\end{equation}
where in both sums, $c$ indexes all cubes in the $A$ sublattice, and $X_c$ ($Z_c$) is defined as the product of Pauli $X$ ($Z$) operators over the vertices of the cube $c$ (see \figref{fig:model}). The model constitutes a stabilizer code Hamiltonian; \cite{Gottesman97} i.e. it is a sum of commuting frustration-free products of Pauli operators, and hence is exactly solvable.

Although there is exactly one Hamiltonian term per qubit, when periodic boundary conditions are imposed, these terms collectively satisfy certain relations which result in a non-trivial ground state degeneracy (GSD). (Note that all three dimensions of the lattice must be even in order for the checkerboard sublattice structure to exist under periodic boundary conditions.) In particular, for each $xy$, $yz$, and $xz$ layer of elementary cubes $L$, we have the following relation:
\begin{equation}
    \prod_{c\in L\cap A}X_c=1,
\end{equation}
and likewise for $Z_c$. For a lattice of size $2L_x\times 2L_y\times 2L_z$, there are thus $4(L_x+L_y+L_z)$ such relations, of which 6 are generated by the remaining relations and hence are redundant.\cite{Sagar16} The GSD therefore obeys the formula
\begin{equation}
    \log_2 \textrm{GSD}=4L_x+4L_y+4L_z-6.
    \label{eqn:GSD}
\end{equation}
A simple observation is that the number of logical qubits (i.e. $\log_2 \textrm{GSD}$) is exactly double that of the X-cube model defined on an $L_x\times L_y\times L_z$ size lattice, which has a code space of $2L_x+2L_y+2L_z-3$ qubits. The characteristic sub-extensive scaling of the GSD can be understood in terms of the renormalization group (RG) transformation discussed in the next section. Therein, two toric code layers are added in order to increase the system size by 2 lattice spacings in one direction, corresponding to an increase in GSD by a factor of 16. 

The logical operators of the model, which map between ground states, correspond to processes in which particle-antiparticle pairs are created out of the vacuum, wound around the spatial manifold, and then annihilated. A salient feature of the model is that these fractional excitations exist within a hierarchy of subdimensional mobility: \textit{planons} are free to move within a plane but cannot leave the plane; \textit{lineons} can move freely along a straight line; whereas \textit{fractons} are fully immobile and cannot be moved whatsoever without creating additional excitations. Moreover, the model has a simple self-duality realized by Hadamard rotation, which is reflected naturally in the particle content. The full structure of excitations is examined more closely in \secref{sec:excitations}.

\section{Renormalization group transformation}

\label{sec:RG}

\begin{figure}
    \centering
    \includegraphics[width=5.66cm]{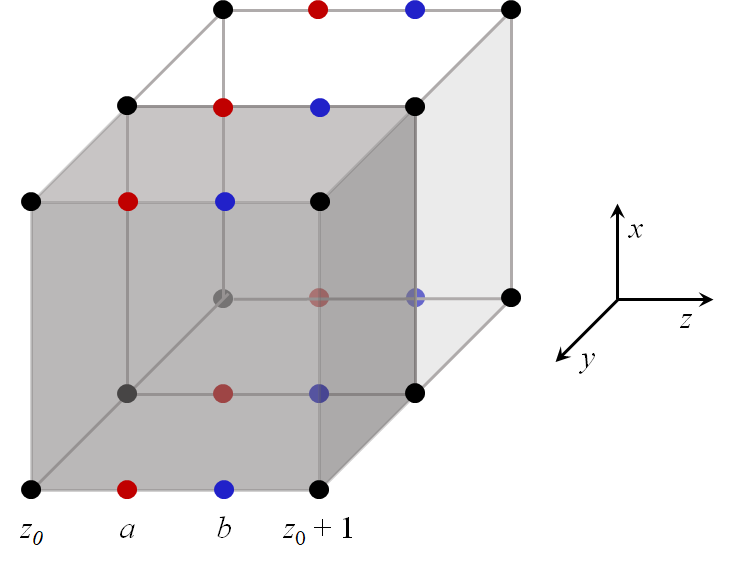}
    \caption{Qubits involved in the RG transformation for the checkerboard model. A single unit cell of the original $2L_x\times 2L_y\times 2L_z$ cubic lattice is depicted here. The black qubits belong to the original checkerboard model. The red and blue qubits comprise the toric code bilayer used as an entanglement resource in the RG procedure and are placed at the vertices of square lattices which are respectively embedded in the $z=a$ and $z=b$ planes. The shaded cube belongs to the $A$ sublattice of the checkerboard bipartition.}
    \label{fig:qubits}
\end{figure}

In this section, we discuss the RG transformation for the checkerboard model, which utilizes toric code bilayers as 2D resources of long-range entanglement. The existence of this transformation establishes the checkerboard model as a fixed-point representative of a foliated fracton phase. The procedure presented here can be compared to the corresponding procedure for the X-cube model\cite{3manifolds}, which uses single toric code layers as 2D resource states. To realize the RG transformation, we construct a local unitary operator $S$ which sews a single toric code bilayer ground state (i.e. two copies of the toric code) into a $2L_x\times 2L_y\times 2L_z$ checkerboard ground state to yield a $2L_x\times 2L_y\times 2\left(L_z+1\right)$ checkerboard ground state. (Since all lattice dimensions must be even, this is the minimal re-sizing allowed.) Arbitrary re-scaling of the model may then be achieved by reversing or iterating this transformation.

\begin{figure}
    \centering
    \includegraphics[width=7.5cm]{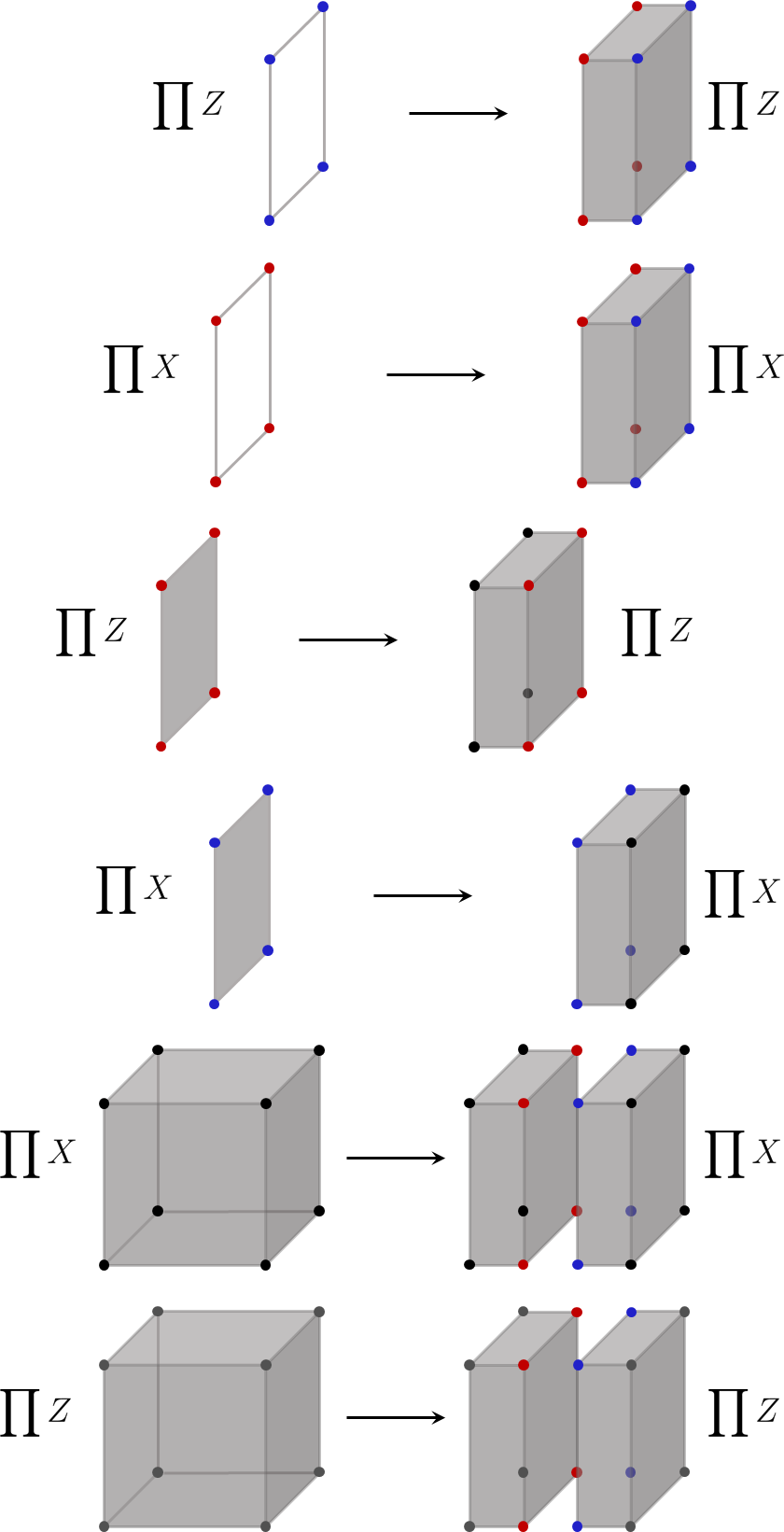}
    \caption{Action of the local unitary $S$ on the stabilizer generators of the composite ground state $\ket{\psi_\mathrm{CB}}\otimes\ket{\psi_{\mathrm{TC}}^a}\otimes\ket{\psi_{\mathrm{TC}}^b}$. Here $\prod X$ $\left(\prod Z\right)$ denotes the product of Pauli $X$ ($Z$) operators over all depicted qubits. On the left side, the shaded cells correspond to the original $A$ sublattice, whereas on the right side shaded cells correspond to the enlarged $A$ sublattice.}
    \label{fig:stabilizers}
\end{figure}

To describe the exact transformation, it is helpful to refer to \figref{fig:qubits}. We label vertices of the original lattice by integrals vectors $(x,y,z)$ where $x=1,2,\ldots,2L_x$ and equivalently for $y$ and $z$. We then consider the tensor product $\ket{\psi_\mathrm{CB}}\otimes\ket{\psi_{\mathrm{TC}}^a}\otimes\ket{\psi_{\mathrm{TC}}^b}$ of the $2L_x\times 2L_y\times 2L_z$ checkerboard ground state $\ket{\psi_\mathrm{CB}}$ with a toric code bilayer ground state $\ket{\psi_{\mathrm{TC}}^a}\otimes\ket{\psi_{\mathrm{TC}}^b}$ living on augmenting $z=a$ and $z=b$ planes lying between the original $z=z_0$ and $z=z_0+1$ lattice layers ($z_0< a< b<z_0+1$). The states $\ket{\psi_{\mathrm{TC}}^a}$ and $\ket{\psi_{\mathrm{TC}}^b}$ are defined as ground states of Hamiltonians $H_\mathrm{TC}^a$ and $H_\mathrm{TC}^b$ on square lattices commensurate with the original cubic lattice. The toric code bilayer qubits, in addition to the original checkerboard model qubits, therefore lie at the vertices of an enlarged $2L_x\times 2L_y\times 2\left(L_z+1\right)$ cubic lattice. $H_\mathrm{TC}^a$ and $H_\mathrm{TC}^b$ are defined as
\begin{equation}
    \begin{split}
        H_\mathrm{TC}^a=-\sum_{p\in A}Z_p-\sum_{p\in B}X_p \\
        H_\mathrm{TC}^b=-\sum_{p\in A}X_p-\sum_{p\in B}Z_p
    \end{split}
\end{equation}
where $p$ runs over all plaquettes in the $A$ or $B$ sublattice and $X_p$ ($Z_p$) is the product of Pauli $X$ ($Z$) operators over the vertices of plaquette $p$. A plaquette $p$ is in sublattice $A$ ($B$) if it is contained within an $A$ ($B$) sublattice cube in the original $2L_x\times 2L_y\times 2L_z$ checkerboard lattice.
(These Hamiltonians are identical to Kitaev's toric code,\cite{KitaevToricCode} except that the underlying square lattice is equivalent to the medial lattice of the square lattice in Kitaev's construction.) This information is summarized on the left hand side of \figref{fig:stabilizers}, which depicts the stabilizer generators of the composite state $\ket{\psi_\mathrm{CB}}\otimes\ket{\psi_{\mathrm{TC}}^a}\otimes\ket{\psi_{\mathrm{TC}}^b}$.

\begin{figure}
    \centering
    \includegraphics[width=4.1cm]{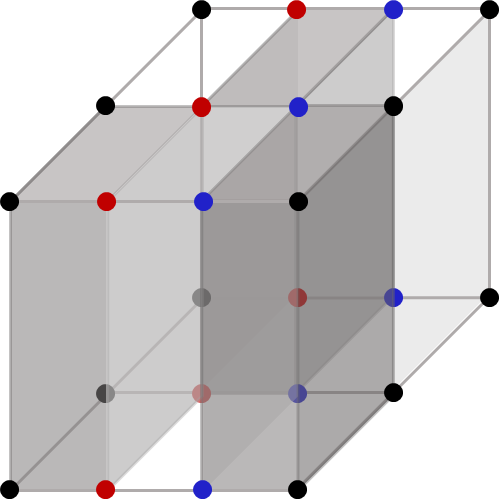}
    \caption{Modified checkerboard sublattice structure after the red and blue qubit layers have been incorporated into the model via the RG transformation. The new $A$ sublattice corresponds to the shaded cells.}
    \label{fig:qubits2}
\end{figure}

To complete the RG procedure, we apply a local unitary operator $S$ in order to yield the enlarged checkerboard ground state $\ket{\psi_\mathrm{CB}}'=S\left(\ket{\psi_\mathrm{CB}}\otimes\ket{\psi_\mathrm{TC}^a}\otimes\ket{\psi_\mathrm{TC}^b}\right)$. Here,
\begin{equation}
    S=\prod_{(x,y)}\mathrm{CX}^{(x,y,a)}_{(x,y,b)}\prod_{(x,y)}\mathrm{CX}^{(x,y,z_0)}_{(x,y,a)}\mathrm{CX}^{(x,y,b)}_{(x,y,z_0+1)}
    \label{eqn:S}
\end{equation}
where $\prod_{(x,y)}=\prod_{x=1}^{2L_x}\prod_{y=1}^{2L_y}$ and $\mathrm{CX}^{(x,y,a)}_{(x,y,b)}$ is defined as the controlled X (i.e. controlled NOT) quantum gate with control qubit $(x,y,a)$ and target qubit $(x,y,b)$. Note that $\mathrm{CX}^{(x,y,z_0)}_{(x,y,a)}$ and $\mathrm{CX}^{(x,y,b)}_{(x,y,z_0+1)}$ commute with one another but not with $\mathrm{CX}^{(x,y,a)}_{(x,y,b)}$. To see that $S$ correctly maps the composite tensor product state to the enlarged checkerboard ground state $\ket{\psi_\mathrm{CB}}'$ one can examine the conjugate action of $S$ on the original stabilizer generators. This is shown graphically in \figref{fig:stabilizers}, recalling that CX acts by conjugation as
\begin{equation}
    \begin{split}
    ZI\to ZI \qquad\qquad IZ\leftrightarrow ZZ \\
    XI\leftrightarrow XX \qquad\qquad IX \to IX.
    \end{split}
\end{equation}
In particular,
\begin{equation}
    S\left(H+H^a_\mathrm{TC}+H^b_\mathrm{TC}\right)S^\dagger\cong H'    
\end{equation}
where $H$ is the original Hamiltonian and $H'$ is the enlarged $2L_x\times 2L_y\times2(L_z+1)$ Hamiltonian, and the $\cong$ operator denotes that the two operators have identical ground spaces. The enlarged $A$ sublattice is depicted in \figref{fig:qubits2}.

\section{General three-manifolds}

\label{sec:manifolds}

In this section, we employ the notion of singular compact total foliation (SCTF), discussed also in \refcite{3manifolds}, to generalize the checkerboard model to compact 3-manifolds other than the 3-torus. An SCTF is a discrete sample of compact leaves of three transversely intersecting (possibly singular) two-dimensional foliations of a 3-manifold $M$, labelled $x$, $y$, and $z$ respectively. For example, the $xy$, $yz$, and $xz$ planes of a cubic lattice embedded in a three-torus may be viewed as the leaves of an SCTF.

\begin{figure}[htbp]
    \centering
    \includegraphics[width=3.5cm]{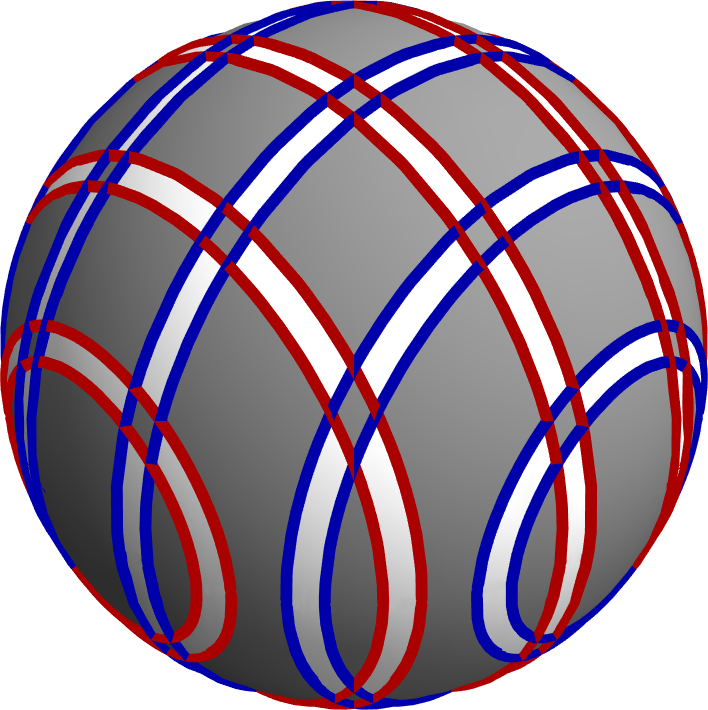}
    \caption{An example of a checkerboard lattice structure embedded in $S^2\times S^1$. Depicted here is an $S^2$ cross-section. The closely-spaced adjacent circles represent bilayers, and the shaded cells belong to the $A$ sublattice.}
    \label{fig:sphere}
\end{figure}

For the checkerboard model, each foliating leaf can be thought of as a bilayer of the underlying lattice of qubits. Thus, to generalize the model we take an SCTF of a 3-manifold $M$ and split each leaf into a bilayer of closely-spaced adjacent parallel leaves. These bilayers constitute a refined SCTF which forms the scaffolding of the embedded lattice. Qubits are placed at triple intersection points of foliating leaves. The elementary 3-cells of the resulting cellulation are then bipartitioned into $A$-$B$ subsets according to the following rule: a 3-cell $c$ belongs to $A$ if it lies within 0 or 2 bilayers, whereas $c$ belongs to $B$ if it lies within 1 or 3 bilayers. See \figref{fig:sphere} for an example of such a structure for the 3-manifold $S^2\times S^1$.

The Hamiltonian of \eqnref{eqn:H} is then readily applied to this generalized checkerboard lattice structure, where in this setting, the $X_c$ ($Z_c$) operator corresponds to products of Pauli $X$ ($Z$) operators over the vertices of 3-cell $c$. As for the checkerboard bipartition of cubic lattice cells, by construction the generalized $A$-$B$ bipartition has the property that all 3-cells of a given partition have an even number of vertices and share an even number of vertices with one another. The Hamiltonian defined in this way is therefore guaranteed to contain mutually commuting terms.

The RG procecedure for the checkerboard model introduced in \secref{sec:RG} can be readily generalized to the model defined via an SCTF on a general 3-manifold. The formula for the GSD in \eqnref{eqn:GSD} therefore generalizes to the form
\begin{equation}
    \log_2\mathrm{GSD}=4g_xL_x+4g_yL_y+4g_zL_z-c
\end{equation}
where $L_\mu$ is the number of leaves in foliation $\mu$, and $g_\mu$ is the genus.\footnote{For non-orientable manifolds, a modified formula is satisfied instead\cite{3manifolds}} The constant $c$ can be computed by using the RG procedure to increasingly coarsen the lattice until the minimal lattice embedding is achieved. We consistently find that $c=2c_\mathrm{XC}$, where $c_\mathrm{XC}$ is the corresponding constant correction to the GSD of the X-cube model defined on the same manifold with the same SCTF (see Table 1 of \refcite{3manifolds}). In all cases the total GSD of the checkerboard model is therefore exactly twice the GSD of the corresponding X-cube model.

\section{Entanglement entropy schemes}

\label{sec:entropy}

Entanglement entropy is a useful way to characterize fracton models.\cite{ShiEntropy,HermeleEntropy,BernevigEntropy,FractonEntanglement} In this section, we briefly discuss the structure of entanglement entropy in the checkerboard model. 

\begin{figure}[htbp]
    \centering
    \includegraphics[width=7cm]{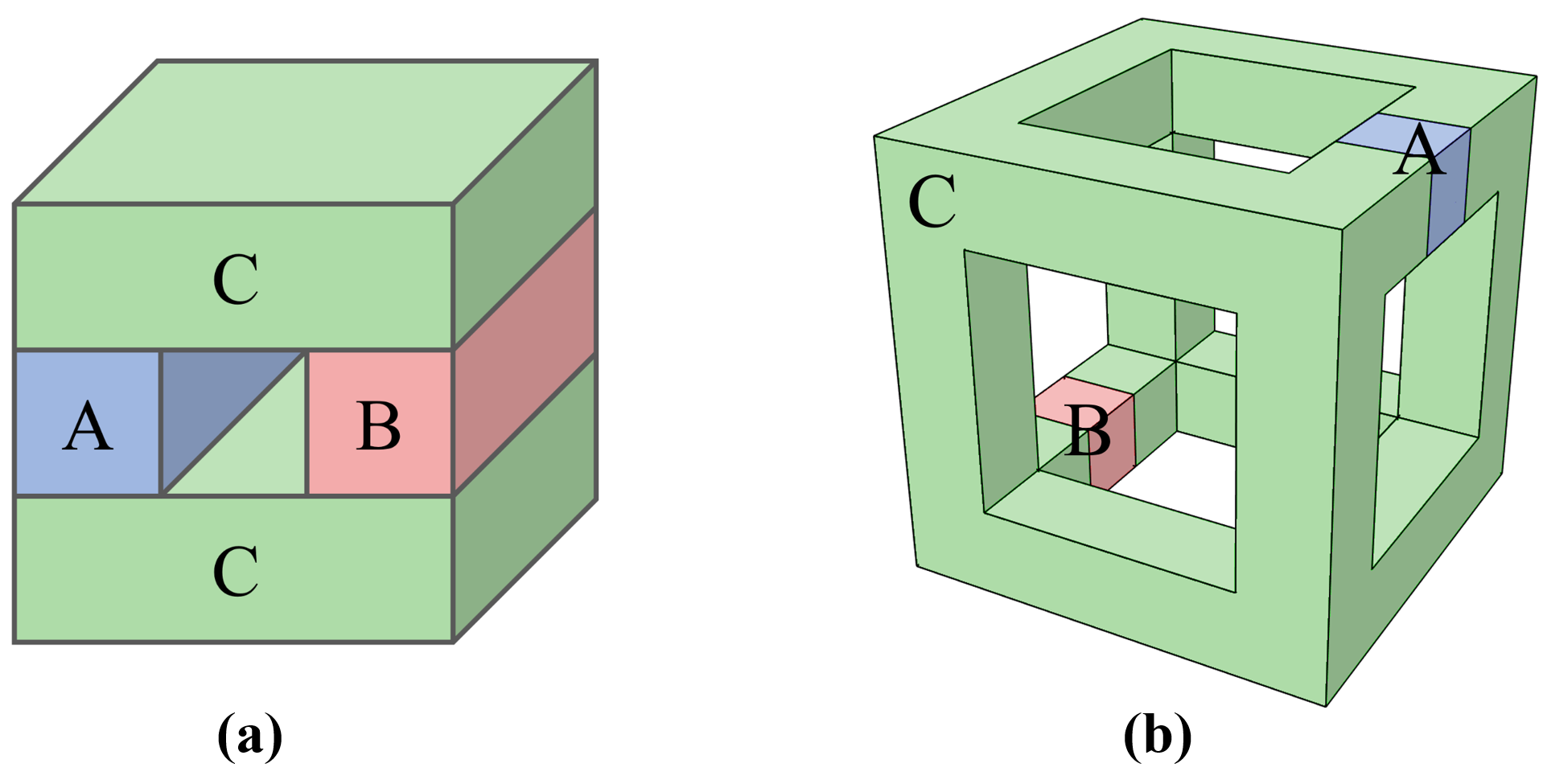}
    \caption{
    {\bf (a)} 3D solid torus $I(A;B|C)$ scheme and {\bf (b)} 3D wire-frame $I(A;B|C)$ scheme. In both cases the regions are contained within an overall cube of side length $L$.
    }
    \label{fig:entanglement}
\end{figure}

Fig.~\ref{fig:entanglement} shows two schemes that can be used to characterize the entanglement structure in the checkerboard model. In both schemes, the quantity to be calculated is
\begin{equation}
I(A;B|C) = S_{AB}+S_{BC}-S_{C}-S_{ABC}
\end{equation}
Applying scheme (a), as proposed in Ref.~\onlinecite{ShiEntropy,HermeleEntropy}, to the checkerboard model, we find that
\begin{equation}
I_a(A;B|C) = 4L+2
\end{equation}
when the overall cubic shape is of linear size $L$ and is aligned with the cubic lattice of the model. $L$ is measured in units of twice the lattice constant of the underlying cubic lattice. As discussed in Ref.~\onlinecite{FractonEntanglement}, the $4L$ term in $I_a$ helps to identify the triple foliation structure revealed by the RG scheme in section~\ref{sec:RG}, since it corresponds to a sum of the topological entanglement entropies of the underlying toric code bilayers.

As discussed in Ref.~\onlinecite{FractonEntanglement}, to characterize foliated topological order beyond the existence of foliation structure, we can use the scheme in Fig.~\ref{fig:entanglement} (b). The foliating layers do not contribute to $I_b(A;B|C)$ in this case and a nonzero $I_b(A;B|C)$ hence represents nontrivial foliated fracton order. Direct calculation shows that
\begin{equation}
I_b(A;B|C) = 2
\end{equation}
for the checkerboard model. This is exactly twice the value calculated for the X-cube model. It is also interesting to note that $I_a$ for the checkerboard model is also exactly twice the value of $I_a$ for the X-cube model, which must be the case in light of the generalized local unitary equivalence demonstrated in \secref{sec:mapping}.

\section{Fractional excitations}

In Ref.~\onlinecite{FractonStatistics}, we propose to characterize fractional excitations in foliated fracton phases using \textit{quotient superselection sectors} and their statistics. In particular, a quotient superselection sector (QSS) is defined as a class of fractional excitations that can be mapped into each other through local operations or by attaching 2D point-like excitations (planons). The universal quasiparticle statistics of a QSS is then captured by applying a set of interferometric operators to the surrounding region of an isolated excitation such that the resulting statistics is the same for excitations in the same QSS. 

Applying these general principles to the checkerboard model, we find that there are six elementary QSS generators, giving rise to a total of $2^6=64$ QSS sectors. It is intructive to take a $2\times 2\times 2$ cell of the underlying cubic lattice as shown in Fig.~\ref{fig:mapping} and to divide the $A$ checkerboard sublattice into four further sublattices $R$, $G$, $B$, and $Y$. The six QSS generators can be taken to be fracton excitations corresponding to a violation of the $X_c$ or $Z_c$ term in the $R$, $G$, and $B$ sublattice cubes respectively, which we label as $f^X_R$, $f^X_G$, $f^X_B$, $f^Z_R$, $f^Z_G$, and $f^Z_B$. Two neighboring fracton excitations in the same sublattice combine into a planon while two neighboring fracton excitations in different sublattices combine into a lineon. Because of this, we could also choose the generating set of QSS to contain two fractons $f^X_R$, $f^Z_R$ and four lineons $f^X_Rf^X_G$, $f^X_Rf^X_B$, $f^Z_Rf^Z_G$, and $f^Z_Rf^Z_B$. As explained in Ref.~\onlinecite{FractonStatistics}, when compared to the X-cube model, we see that this is exactly double the QSS content of the X-cube model.

\begin{figure}[htbp]
\centering
\includegraphics[width=0.48\textwidth]{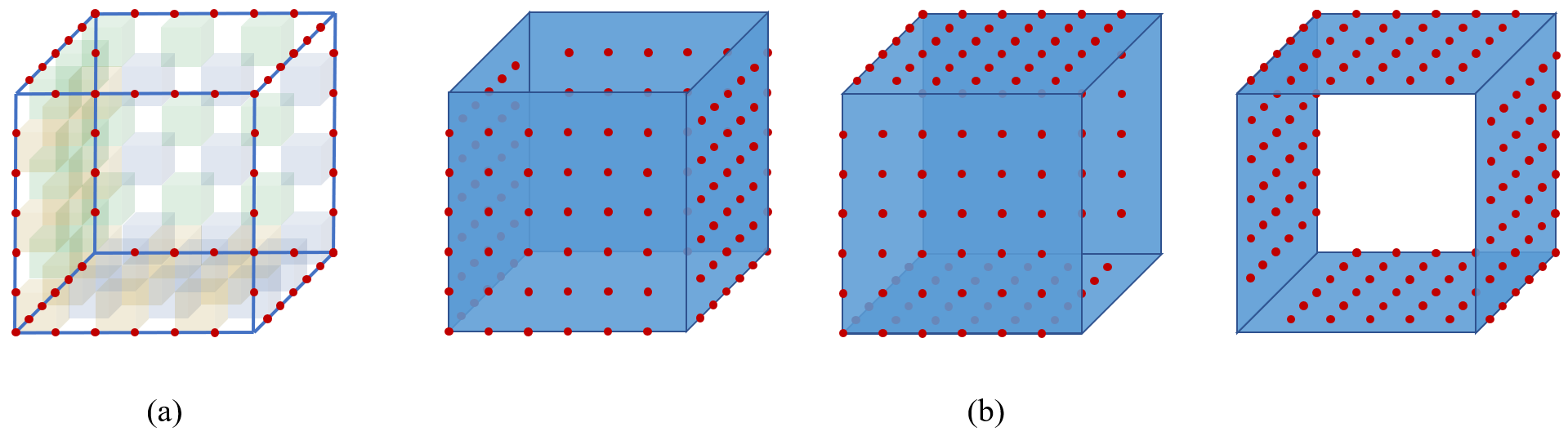}
\caption{Examples of (a) a wireframe operator and (b) membrane operators in the checkerboard model. The operators are tensor products of Pauli $X$ or $Z$ over the red qubits. Shaded cubes belong to the $A$ sublattice.}
\label{fig:CBoperators}
\end{figure}

To detect the quotient charge of an isolated point excitation (i.e. which QSS it belongs to), we can apply interferometric operators as shown in Fig.~\ref{fig:CBoperators}. The operators are tensor products of Pauli $X$ or $Z$ over the red qubits. The wireframe operator can be obtained as a product of all the $X_c$ or $Z_c$ cube operators inside the wireframe. The membrane operators can be obtained as a product of all the cube operators in every other layer inside the overall cube. The number of independent interferometric operators is twice that of the X-cube model and, as shown in Ref.~\onlinecite{FractonStatistics}, there is a mapping between quotient superselection sectors and interferometric operators of the two models which preserves the fusion rules and quasi-particle statistics.

\label{sec:excitations}

\section{Relation to two copies of the X-cube model}

\label{sec:mapping}

In this section, we exhibit an exact local unitary mapping between the checkerboard model ground space on a $2L_x\times 2L_y\times 2L_z$ lattice (denoted $G_{\mathrm{CB}}$) and the ground space of two copies of the X-cube model tensored with product state ancilla qubits on an $L_x\times L_y\times L_z$ lattice (denoted $G_{2\mathrm{XC}}$). The mapping is not a full equivalence of Hamiltonians as it rearranges the energy levels of excitations, but the Hamiltonians are shown to be equivalent as stabilizer codes, and thus have coinciding ground spaces. The X-cube model, as originally discussed in \refcite{Sagar16}, is defined on a cubic lattice with one qubit per edge, and Hamiltonian
\begin{equation}
    H_\mathrm{XC}=-\sum_v\left(A_v^{xy}+A_v^{yz}+A_v^{xz}\right)-\sum_c B_c,
    \label{eqn:Xcube}
\end{equation}
where $v$ runs over all vertices of the lattice and $c$ runs over all elementary cubes of the lattice. The operator $A_v^{xy}$ is defined as the product of Pauli $Z$ operators over the four edges adjacent to vertex $v$ along the $xy$ plane, while $B_c$ is given by the product of Pauli $X$ operators over the edges of the cube $c$.

\begin{figure}
    \centering
    \includegraphics[width=5cm]{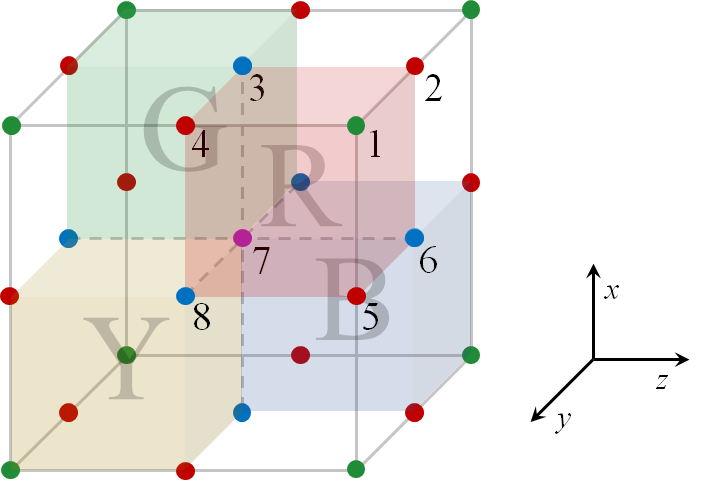}
    \caption{Matching of qubits between the checkerboard model and two copies of the X-cube model tensored with ancilla qubits. A $2\times2\times2$ cell of the checkerboard model cubic lattice is shown here, corresponding to a single unit cell of $\Lambda$, whose vertices lie at the green points. Shaded cubes belong to sublattice $A$ of the checkerboard bipartition. The red and blue qubits located respectively on the direct lattice (solid lines) and dual lattice edges (dashed lines) belong to the two X-cube copies, whereas the green and purple qubits at the vertices and body-center are ancilla degrees of freedom. The numbers label the qubits of a single unit cell of $\Lambda$.}
    \label{fig:mapping}
\end{figure}

To match the degrees of freedom of the two systems, we start with an $L_x\times L_y\times L_z$ cubic lattice whose points are labelled by vectors $(x,y,z)$ and belong to the set $\Lambda$ ($x=1,2,\ldots,L_x$ and equivalently for $y$ and $z$). We then place one set of qubits on the edges of the lattice, corresponding to one copy of the X-cube model with Hamiltonian $H_\mathrm{XC}^1$, and another set of qubits on the edges of the dual lattice (i.e. the plaquettes of the direct lattice), corresponding to the second copy of the X-cube model, whose Hamiltonian $H_\mathrm{XC}^2$ is transformed relative to \eqnref{eqn:Xcube} via a global Hadamard rotation ($X\leftrightarrow Z$). Finally, ancilla qubits are placed at the vertices and body-centers of the lattice, and initialized in $+1$ eigenstates of the Pauli $Z$ and $X$ operators respectively. As shown in \figref{fig:mapping}, all the qubits together constitute a cubic lattice of dimensions $2L_x\times 2L_y\times 2L_z$ and half the lattice spacing of the original model. There are thus 8 qubits in each unit cell of $\Lambda$, which are numbered according to the scheme in \figref{fig:mapping}.

\begin{figure}
    \centering
    \includegraphics[width=8.65cm]{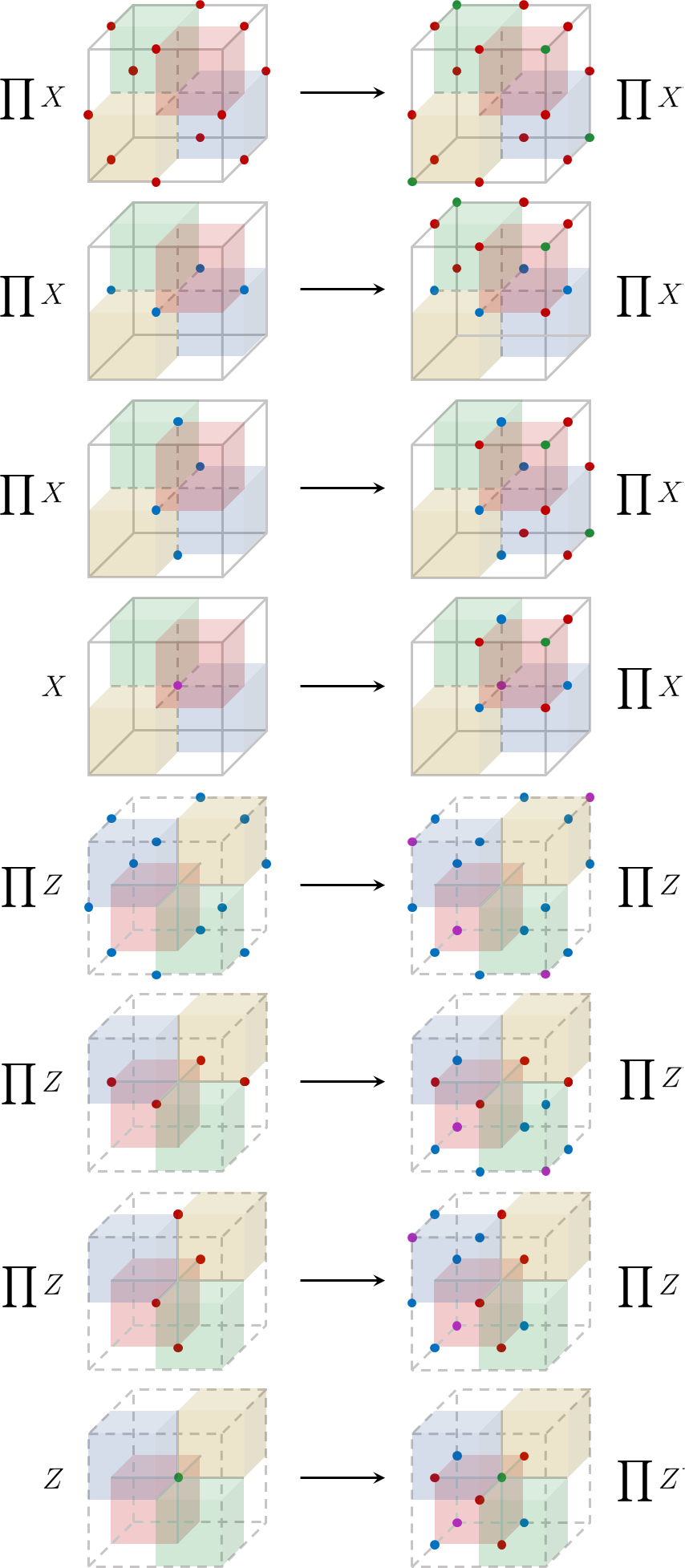}
    \caption{Action of $U$ on the stabilizer generators of $G_{2\mathrm{XC}}$. Here $\prod X$ $\left(\prod Z\right)$ denotes the product of Pauli $X$ ($Z$) operators over all depicted qubits. Solid lines correspond to direct lattice edges, whereas dashed lines correspond to dual lattice edges. From top to bottom, the image terms equate to $X_RX_GX_BX_Y$, $X_RX_G$, $X_RX_B$, $X_R$, $Z_RZ_GZ_BZ_Y$, $Z_RZ_G$, $Z_RZ_B$, and $Z_R$ operators in the checkerboard model (\eqnref{eqn:H}) respectively, where $R$, $G$, $B$, and $Y$ refer to the red, green, blue, and yellow cubes.}
    \label{fig:stabilizers2}
\end{figure}

To demonstrate equivalence of the two ground spaces, consider the local unitary operator $U=U_2U_1$ where
\begin{equation*}
        U_1=\prod_{v\in\Lambda}
        \mathrm{CX}^{v,2}_{v,1}
        \mathrm{CX}^{v,4}_{v,1}
        \mathrm{CX}^{v,5}_{v,1}
        \mathrm{CX}^{v,7}_{v,3}
        \mathrm{CX}^{v,7}_{v,6}
        \mathrm{CX}^{v,7}_{v,8}
\end{equation*}
and
\begin{equation*}
    \begin{gathered}
        U_2=\prod_{v\in\Lambda}
        \mathrm{CX}^{v,7}_{v,1}\times\\
        \mathrm{CX}^{v,3}_{v,2}
        \mathrm{CX}^{v,3}_{v-{\bf\hat{y}},4}
        \mathrm{CX}^{v,6}_{v,2}
        \mathrm{CX}^{v,6}_{v-{\bf\hat{y}},5}
        \mathrm{CX}^{v,8}_{v,4}
        \mathrm{CX}^{v,8}_{v,5}\times\\
        \mathrm{CX}^{v,8}_{v,1}
        \mathrm{CX}^{v+{\bf\hat{y}},3}_{v,1}
        \mathrm{CX}^{v+{\bf\hat{y}},6}_{v,1}
        \mathrm{CX}_{v,2}^{v,7}
        \mathrm{CX}_{v-{\bf\hat{y}},4}^{v,7}
        \mathrm{CX}_{v-{\bf\hat{y}},5}^{v,7}.
    \end{gathered}
\end{equation*}
Here $\mathrm{CX}^{v,a}_{u,b}$ denotes a controlled X gate with control qubit $a$ at point $v\in\Lambda$ and target qubit $b$ at point $u\in\Lambda$. The conjugate action of $U$ on the stabilizer generators of the code space $G_{2\mathrm{XC}}$ is shown graphically in \figref{fig:stabilizers2}. Note that, because two of the three vertex stabilizers generate the third, it is sufficient to consider the action on just two vertex terms. The image stabilizers on the right-hand side are products of stabilizer terms for the checkerboard model, and generate a stabilizer code identical to that of the checkerboard Hamiltonian. In particular,
\begin{equation}
    UH_\mathrm{CB}U^\dagger\cong H^0+H^1_\mathrm{XC}+H^2_\mathrm{XC}
\end{equation}
where $H_\mathrm{CB}$ is the checkerboard Hamiltonian and $H^0$ acts on the ancilla degrees of freedom.

\section{Discussion}
\label{sec:discussion}

In this paper we show that the checkerboard model (first discussed in \refcite{Sagar16}) belongs to a foliated fracton phase, as defined in \refcite{3manifolds}. Moreover, we identify the foliated fracton order in the checkerboard model to be equivalent to that of two copies of the X-cube model (also introduced in \refcite{Sagar16}). This is, in a sense, similar to the equivalence between the 2D color code and two copies of the 2D toric code as conventional topological order.

The existence of such an equivalence is far from obvious as the two models in their original form appear to have significant differences. The checkerboard model has elementary (with minimum energy) lineons whose string operators may anti-commute with each other, which is not the case for the elementary lineons of the X-cube model. Moreover, in the checkerboard model an elementary lineon is the composite of two elementary fractons, which is not the case in the X-cube model. Such differences may seem significant, but they are actually superficial as they depend sensitively on which excitations are considered the `elementary’ ones, which is not a universal property of a phase.

The explicit mapping (\figref{fig:stabilizers}) between the two models allows us to see that an elementary fracton in the checkerboard model is related to a composite fracton in the pair of X-cube models, which is a bound state of elementary X-cube fractons and lineons (along with a possible ancillary bosonic excitation).
The elementary lineon in the checkerboard model, which is a bound state of two elementary fractons, is then related to a composite lineon in the X-cube models, which is a bound state of two composite fractons: i.e. a bound state of fracton dipoles (2D particles) and elementary lineons in the X-cube models.
Because these composite lineons are made of conjugate fracton dipoles and lineons, their string operators may anti-commute, similar to the string operators in the checkerboard model.
This resolves the apparent differences between the checkerboard and pair of X-cube models discussed in the previous paragraph.

While the superficial differences can obscure the intrinsic relation between the fracton orders in different fracton models, by considering their universal properties such as the foliation-free entanglement entropy and fractional statistics, we are able to see clearly the equivalence between the checkerboard model and two copies of the X-cube. Note that the mapping we found between the two models is special in that we only need to add product state ancillas before doing local unitary transformations. In general, if two models have the same foliated fracton universal properties, then to connect them we may need to add two dimensional gapped states as resource before applying local unitary operations. In \refcite{FractonStatistics}, we present such an example (between the X-cube model and the semionic X-cube model). 

With the definition given in \refcite{3manifolds} and the universal properties defined in Refs.~\onlinecite{FractonEntanglement} and  \onlinecite{FractonStatistics}, we have a established a useful set of tools to study foliated fracton order. It would be interesting to explore various other models and identify different types of foliated fracton order, from which a more systematic understanding of the phenomenon may be established.

\begin{acknowledgements}
We are grateful for inspiring discussions with Abhinav Prem. W.S. and X.C. are supported by the National Science Foundation
under award number DMR-1654340 and the Institute for Quantum Information and Matter
at Caltech. X.C. is also supported by the Alfred P. Sloan research fellowship and the Walter Burke Institute for Theoretical Physics at Caltech.
K.S. is grateful for support from the NSERC of Canada, the Center for Quantum Materials at the University of Toronto, and the Walter Burke Institute for Theoretical Physics at Caltech.
\end{acknowledgements}

\end{document}